\documentclass[%
superscriptaddress,
twocolumn,
aps,
prl,
]{revtex4-2}

\usepackage{graphicx}% Include figure files
\usepackage{dcolumn}% Align table columns on decimal point
\usepackage{bm}% bold math
\usepackage[normalem]{ulem}
\usepackage{xcolor} % colored text
\usepackage{hyphenat}
\usepackage{float}
\usepackage{amssymb}
\usepackage{amsmath}

\usepackage{caption}
\usepackage{subcaption}
\usepackage{ragged2e} % Import the ragged2e package
\usepackage[colorlinks,linkcolor=blue,anchorcolor=pink,filecolor=green,menucolor=yellow,citecolor=blue,urlcolor=blue,bookmarks=false,hypertexnames=true]{hyperref}

\usepackage{orcidlink} % CTAN package

\newcommand{\be}{\begin{equation}}
\newcommand{\ee}{\end{equation}}

\newcommand{\tauR}{\tau_\mathrm{R}}
\newcommand{\DR}{D_\mathrm{R}}
\newcommand{\DT}{D_\mathrm{T}}
\newcommand{\vmax}{v_0}
\newcommand{\vmin}{v_1}
\newcommand{\RQS}{R_{\mathrm{QS}}}
\newcommand{\AQS}{\pi R_{\mathrm{QS}}^2}
\newcommand{\kin}{k_\mathrm{in}}
\newcommand{\kout}{k_\mathrm{out}}
\newcommand{\fc}{f_\mathrm{c}}

\newcommand{\rhog}{\rho_{\rm g}}
\newcommand{\rhoc}{\rho_{\rm c}}

\newcommand{\rhoglobal}{\rho_{0}}

\begin{document}

\preprint{APS/123-QED}

\title{
How Quorum Sensing Shapes Clustering in Active Matter
%Active Matter with Quorum Sensing Exhibits Reentrant Clustering\\
%Quorum Sensing Induces Reentrant Clustering in Active Matter\\
%Active Particles with Quorum Sensing Exhibit Reentrant Clustering\\
%Quorum Sensing Induces Reentrant Clustering in Active Particles\\
% Reentrant Clustering in Active Particles Induced by Quorum Sensing\\
% How Quorum Sensing Affects Motility-Induced Clustering\\
% Active Particles Interacting via Contact Forces and Quorum Sensing Motility Regulation\\
% Clustering of Active Matter with Steric and Quorum Sensing Interactions\\
% Self-propelled disks with quorum-sensing motility regulation\\
% Active particles interacting via chemical and mechanical motility regulation
}
\author{L. de Souza\,\orcidlink{0000-0003-0941-3568}}
\email{lucas.gabrielsouza@fisica.ufrn.br}
\affiliation{Departamento de {Física} {Teórica} e Experimental, Universidade Federal do Rio Grande do Norte. Natal 59078-970, Brazil}
\affiliation{Institut für Theoretische Physik, University of Göttingen, Friedrich-Hund-Platz 1, 37077 Göttingen, Germany}
\author{E.~F. Teixeira\,\orcidlink{0000-0001-5409-6405}}
\email{teixeiraemanuel9@gmail.com}
\affiliation{Instituto de {Física}, Universidade Federal do Rio Grande do Sul. Porto Alegre 91501-970, Brazil}
\author{G.~M. Viswanathan\,\orcidlink{0000-0002-2301-5593}}
\email{gandhi@fisica.ufrn.br}
\affiliation{Departamento de {Física} {Teórica} e Experimental, Universidade Federal do Rio Grande do Norte. Natal 59078-970, Brazil}
\affiliation{National Institute of Science and Technology of Complex Systems. Rio de Janeiro 22290-180, Brazil}
\author{P. Sollich\,\orcidlink{0000-0003-0169-7893}}
\email{peter.sollich@uni-goettingen.de}
\affiliation{Institut für Theoretische Physik, University of Göttingen, Friedrich-Hund-Platz 1, 37077 Göttingen, Germany}%
\affiliation{Department of Mathematics, {King’s} College London. London WC2R 2LS, UK}
\author{P.~de Castro\,\orcidlink{0000-0003-3199-926X}}
\email{pablo.castro@unesp.br}
\affiliation{ICTP South American Institute for Fundamental Research, Instituto de {Física}
{Teórica}, Universidade Estadual Paulista– UNESP, São Paulo, Brazil}
\date{\today}

\begin{abstract}
%\new
{
Self-propelled particles undergoing persistent motion can accumulate either through excluded-volume interactions or through quorum sensing, where self-propulsion decreases at high local density. Using kinetic balance theory and simulations, we show that the interplay of these two mechanisms produces a reentrant, non-monotonic behavior in which clustering passes through a pronounced minimum as quorum-sensing strength or persistence time varies. Beyond a threshold quorum sensing strength, we find long-lived transient states that retain memory of initial conditions, including kinetically arrested active gels. Although quorum sensing can mimic attractive interactions, it also acts strongly in dilute regions, producing an effective cluster bistability that is captured by our theory. Our results explain collective states observed experimentally in synthetic and biological active systems.}
\end{abstract}

\maketitle

{\it Introduction}---Active matter refers to nonequilibrium systems composed of large ensembles of self-propelled entities~\cite{teixeira2021single,ourique2022modelling,van2024soft}, such as bacteria~\cite{Sokolov2007Apr}, cell tissues~\cite{teixeira2025collective,PhysRevLett.134.138401}, animal flocks~\cite{Ballerini2008Jan}, and colloids~\cite{Buttinoni2013Jun}. When the directions of self-propulsion fluctuate slowly, the interplay of repulsive contact forces and persistent motion can generate a clustering phenomenon known as motility-induced phase separation (MIPS)~\cite{Henkes2011Oct,rojas2023wetting,Fily2012Jun,Forget2022Dec}. This minimal set of ingredients explains the behavior of many synthetic active systems~\cite{Buttinoni2013Jun,palacci2013living}.
% However, many species, including microorganisms, often interact via chemical signals which trigger motility regulation, in a mechanism called quorum sensing (QS)~\cite{Liu2013Jul,Liu2019Jun,Zhao2023Sep,Liu2019Jun,Anderson2022Nov}. \cmPa{Are refs appropriate?}
% Experiments with mussels, bacteria and fire ants indicate that, at moderate densities, MIPS may not be the only possible phenomenon and may even be suppressed by QS~\cite{Liu2013Jul,DalCo2020Mar,vanGestel2021Apr,Anderson2022Nov}.
% Recent experiments with synthetic colloidal rods that mimic quorum-sensing mechanisms revealed a novel phase coexistence mechanism caused by the interplay of quorum sensing and steric contact interactions~\cite{Lefranc2025Aug}.\cmPa{a bit vague} For simplicity, QS is often overlooked in models with steric interactions~\cite{bechinger2016active}.
% %, particularly when the system has moderate density.
% Conversely, when QS is taken into account, steric interactions are typically neglected~\cite{Dinelli2023Nov}, rendering current QS models useful only at low densities~\cite{Curatolo2020Nov}.
However, organisms often interact via chemical signals that regulate self-propulsion through a mechanism called quorum sensing (QS)~\cite{Liu2013Jul,Liu2019Jun,Zhao2023Sep,Anderson2022Nov,Curatolo2020Nov}, which can help survival~\cite{daniels2004quorum}. 
Experiments with bacteria, ants, and mussels show that, at moderate densities, QS can supress contact-force MIPS~\cite{Liu2013Jul,DalCo2020Mar,vanGestel2021Apr,Anderson2022Nov}. 
Recent experiments with synthetic QS rods~\cite{Lefranc2025Aug} and simulations of QS disks~\cite{Nguyen2025Jul} showed that the interplay of QS and steric repulsion leads to arrested dynamics and shapes clustering. For simplicity, most models include either steric forces or QS, but rarely both~\cite{Martinez_Garcia_2015Jul,bechinger2016active,ridgway2023motility}, which limits their applicability.

A common effective description of QS incorporates motility regulation via a self-propulsion speed that decreases with the local density of individuals, without explicitly modeling signaling molecules~\cite{Dinelli2023Nov}. In coarse-grained approaches, this type of QS motility reduction can lead to an effective attraction~\cite{Tailleur2008May,Cates2015Mar}. For truly attractive active particles, persistent motion can counterbalance attraction, leading instead to homogeneous states~\cite{Redner2013Jul}. This competition produces a \textit{reentrant} clustering phase diagram: as activity is varied, clustering disappears and then reappears~\cite{Pu2017May,Sarkar2021Feb,Gutierrez2021Oct,Bhowmick2025Jun}. 
How such reentrant behavior manifests in QS particles remains unclear. 
% Moreover, despite superficial similarities, QS motility reduction and true attraction may generate distinct dynamics~\cite{Lefranc2025Aug}.
% \new{In particular, Ref.~\cite{Nguyen2025Jul} justifies that QS and steric interactions can induce metastable phases and shows that their competition can hinder the growth of cluster density.}
More generally, although a few works have included both QS and steric interactions~\cite{Bauerle2018Aug,Fischer2020Jan,Jose2021Mar,Lefranc2025Aug,Nguyen2025Jul}, the clustering behavior of these ubiquitous systems is an open problem.

%
% In previous models involving QS, steric interaction and persistent motion~\cite{Bauerle2018Aug,Fischer2020Jan,Jose2021Mar,Lefranc2025Aug,Nguyen2025Jul}, QS and steric interactions were can lead to novel behavior.
% Those provided a more realistic perspective on biological and bio-inspired synthetic systems.
%

%
% In this work, we show that active particles interacting via both steric repulsion and quorum-sensing motility
% reduction exhibit a reentrant clustering behavior driven by a novel class of active gels.
% %
% Moreover, we show that, while
% quorum sensing can mimic passive attraction, it also produces strong and previously unexplored
% effects in dilute regions, fundamentally altering the phase diagram. Our analysis involves large-scale simulations of active Brownian particles with QS motility reduction via a density-dependent self-propulsion speed. Also, we develop a kinetic balance theory for the clustering degree, from which we predict two effective equilibria depending on initial conditions as a result of kinetically-arrested states. This initial-condition dependent behavior, confirmed via simulations, has strong resemblance with experiments in mussel beds~\cite{vandeKoppel2008Oct,Liu2013Jul}. Our results connect phenomena
% observed in synthetic and biological systems and show that quorum sensing, historically overlooked
% in modeling, reshapes self-organization, opening new avenues in the physics of active matter.
Here, we demonstrate that combining steric repulsion with quorum-sensing motility reduction leads to a reentrant clustering behavior, where a previously unexplored class of active kinetically-arrested \textit{gels} arises~\cite{Redner2013Jul}. These are disordered solid-like percolating networks~\cite{Segre2001Jun,Prost2015Feb} which are induced here by QS. Furthermore, QS not only mimics attractive potentials but also induces pronounced effects in dilute regions, which qualitatively transform the phase diagram. Beyond numerical simulations, we derive a kinetic balance theory for the clustering degree by modifying the approach of Ref.~\cite{Redner2013Jul}. The proposed theory predicts that, depending on the initial condition, different effective equilibria arise. This prediction is confirmed in simulations and resembles behaviors observed in mussel-bed experiments~\cite{vandeKoppel2008Oct,Liu2013Jul}. Taken together, our results bridge synthetic and biological realizations of active matter, demonstrating how quorum sensing, often neglected in modeling efforts, plays a central role in shaping collective organization.
% %short-range
% The system has a moderate packing fraction of ${\phi=0.4}$ that increases the number of particle collisions.
% %
% We find that persistent motion and QS compete to drive aggregation, leading to non-monotonic clustering behavior.
% %
%In the adequate region of parameter space, we find that persistent motion and motility reduction compete to drive aggregation, resulting in a non-monotonic clustering behavior and an increase in the intensity of either aggregation mechanism.
\begin{figure*}
\includegraphics[width=\linewidth]{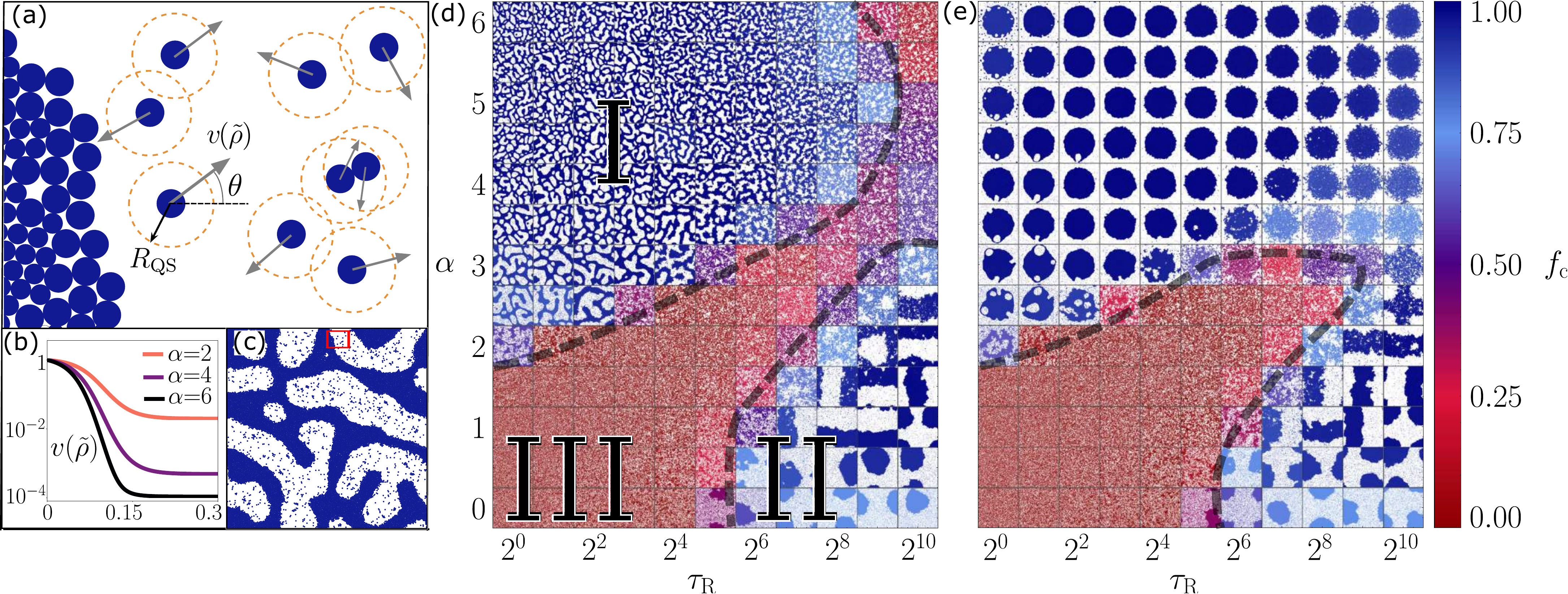}
    \caption{\justifying
    Quorum sensing gives rise to active gels, reentrant clustering, and long-lived transients.
    (a) Disk-like particles form an arrested cluster and a dilute phase. 
    Each particle self-propels in a direction defined by $\theta$ (gray arrows), with a speed $v$ that decreases with the weighted density $\tilde{\rho}$ of particles within a radius $\RQS$. 
    (b) $v(\tilde{\rho})$ [Eq.~\eqref{eq:speedQS}] drops from one plateau to another
    %across a transition localized around ${\bar{\rho} = 0.1}$, with width ${\delta \rho = 0.05}$.
    % $\alpha$ is the quorum-sensing ``strength''.
    (c) Zoom-out of panel (a) reveals a gel pattern caused by low directional persistence and speeds. 
    Here, ${\tauR=4}$, ${\alpha=3}$, $\phi=0.4$, ${N=2\times 10^4}$, and ${t=2 \times 10^6}$. 
    (d) Phase diagram in the $\tau_R$–$\alpha$ plane, based on snapshots at ${t = 2 \times 10^6}$ from a homogeneous initial condition. 
    The color bar shows $f_c$, 
    % the fraction of particles in clusters with at least $10^3$ particles, 
    averaged between ${t = 1.7 \times 10^6}$ and ${2\times 10^6}$. 
    A reentrant $f_c$ reveals three regions. 
    Region I shows gels as in panel (c). 
    Region II features motility-induced phase separation due to contact forces.
    In Region III, slowed particles have just enough persistence to escape without forming clusters. 
    (e) Same as panel (d), but the initial condition is a single cluster with inward-pointing particles at the interface. 
    There is a long-lived transient due to the memory of the initial condition. 
    The diagram in panel (e) converges slowly to that in panel (d); see Supplementary Material~\cite{SM}.
    }
    \label{fig:phdiag_tau_alpha}
\end{figure*}

{\it Model}---We consider $N$ active Brownian particles (ABPs) in two dimensions with periodic boundary conditions. Their dynamics is
\be \label{eq:posdyn}
\dot{\mathbf{r}}_i(t) = v[ \tilde{\rho}(\mathbf{r}_i) ] \, {\bf \hat{e}}_i + \mu \displaystyle\sum_{j \neq i} \mathbf{F}_{i j} + \sqrt{2 \DT} \,\boldsymbol{\xi}_i(t),
\ee
\be \label{eq:orientdyn}
\dot{\theta}_i(t) = \sqrt{2 \DR} \, \eta_i(t),
\ee
where ${\bf r}_i$ is the position, ${{\bf \hat{e}}_i = (\cos \theta_i, \sin \theta_i)}$ is the self-propulsion direction (or polarity), $\mu$ is the mobility, $\xi_{i,x}$, $\xi_{i,y}$, and $\eta_i$ are independent unit Gaussian white noises, and $\DT$ and $\DR$ are the translational and rotational noise strengths, respectively, treated as independent parameters. We assume strong elastic repulsive contact forces
$
{\mathbf{F}_{i j} = k \, \left( d_{ij} - r_{i j} \right) \Theta\left( d_{ij} - r_{i j} \right) \hat{\mathbf{r}}_{i j}},
$
where $d_{ij} = R_i + R_j$ and $R_i$ are the particle radii. To avoid artificial crystallization, each particle is randomly assigned one of two radii. The average diameter is ${\sigma = 1}$ and the ratio between radii is $1.4$~\cite{Tong2019Dec}. We work at a moderate packing fraction $\phi = 0.4$ (that is, number density $\rhoglobal \approx 0.49$), with ${N=2\times 10^4}$ particles.
% , well below the geometric percolation threshold

The quorum-sensing motility regulation is modeled by a density-dependent self-propulsion speed~\cite{Dinelli2023Nov}
\be \label{eq:speedQS}
v(\tilde{\rho}) = \vmin + (\vmax-\vmin) \; \exp \left( -\alpha \, {\cal S}(\tilde{\rho} ) \right),
\ee
where ${\tilde \rho}$ is a particle density calculated within a QS range, $\RQS$, around the focal particle; see Fig.~\ref{fig:phdiag_tau_alpha}a and Movie~S1 in the Supplementary Material (SM)~\cite{SM}. Each particle inside the QS range contributes to $\tilde{\rho}$ with a weight that depends on its distance, as specified by a kernel function given in the SM~\cite{SM}. Moreover, $\alpha$ is dubbed QS strength, $\vmax$ is the self-propulsion speed without QS and $\vmin$ for infinite $\alpha$, with ${\vmax>\vmin}$. Also, ${\cal S}(\tilde{\rho}) = \tanh\left(\frac{\tilde{\rho} - \bar{\rho}}{\delta \rho}\right) + \tanh\left(\frac{\bar{\rho}}{\delta \rho}\right)$ is a sigmoid function~\cite{Curatolo2020Nov,Duan2023Oct,Dinelli2023Nov}. This choice for $v(\tilde{\rho})$ resembles activation and saturation effects of chemical receptors in microorganisms. Around ${\tilde \rho}={\bar \rho}$, $v({\tilde \rho})$ undergoes a crossover between two limiting values; see Fig.~\ref{fig:phdiag_tau_alpha}b. How strongly $v({\tilde \rho})$ decays depends on the QS strength $\alpha$ and the decay width $\delta \rho$. 
%We fix ${{\bar \rho}=0.1}$, $\delta \rho=0.05$, $\RQS=1.5\sigma$, $\vmax=1$, and $\vmin=10^{-4}$.
%
The polarity persistence time is ${\tauR\equiv\DR^{-1}}$. We vary $\alpha$ and $\tauR$ and fix the other parameters. For additional simulation details and fixed parameter values, see SM~\cite{SM}. In particular, we employ $\DT\ll \vmax^2\tauR $ and $\vmin\ll\vmax$ as observed for real microorganisms \cite{daniels2004quorum,villa2020run,Curatolo2020Nov}. Nevertheless, $\DT$ and $\vmin$ still serve to eventually break kinetic arrest in our long simulations. 

{\it Reentrant behavior and effective equilibria}---First, we employ a homogeneous initial condition. To estimate clustering, we calculate the fraction of particles in clusters, $\fc$. We consider two particles to be ``connected'' when $r_{ij}<1.1 d_{ij}$. A cluster is defined as any connected group containing more than $5\%$ of all particles. Figure~\ref{fig:phdiag_tau_alpha}d shows a phase diagram in the $\tau_R$–$\alpha$ plane based on snapshots at a late time. Each snapshot is colored according to $\fc$. The phase diagram displays three regions: two associated with strong clustering ($\fc \approx 1$) and, in between them, a region characterized by nearly homogeneous states ($\fc \approx 0$). Owing to this non-monotonic $\fc$, the phase diagram is said to be reentrant. In fact, two distinct reentrant behaviors appear: one upon varying $\tau_R$ and another upon varying $\alpha$.

With the aim of understanding the origin of this phase diagram, let us consider what happens to a particle at the edge of a cluster.
For low $\tau_R$ and high $\alpha$, even if the particle is oriented away from the cluster, its self-propulsion speed is too small to allow for the particle to escape before it reorients back toward the cluster. As a result, particles undergo a slow coarsening process, phase-separating into a gas and a kinetically arrested gel~\cite{Segre2001Jun}, where particles are almost static; see Fig.~\ref{fig:phdiag_tau_alpha}c and Movie~S2 in the SM~\cite{SM}. 
This active gel state arises in Region~I of Fig.~\ref{fig:phdiag_tau_alpha}d. Whereas gelation is typically driven by attractive interactions~\cite{Redner2013Jul}, here it emerges through a QS-induced route. 
By contrast, a distinct route to active gelation without attraction has been reported for particles with non-convex shapes~\cite{Merrigan2020Mar}.
%Due to the particles' small speed, these clusters attach and percolate (see Supplementary Material~\cite{SM} Movie~S1), leading to patterns as in Fig.~\ref{fig:phdiag_tau_alpha}c.

In the opposite limit of high $\tau_R$ and low $\alpha$, we observe conventional MIPS driven by persistent motion and steric contact repulsion~\cite{Redner2013Jan}, as shown in Region~II. The critical value of $\tau_R$ for the transition to Region~II increases with $\alpha$, since stronger QS reduces the propulsion speed and thus requires higher persistence to sustain contact-force MIPS. In Region~III, particles escape QS but fail to undergo MIPS, resulting in a nearly homogeneous state. The III--II transition was first reported for QS disks in Ref.~\cite{Jose2021Mar}, but for their parameters Region~I was not observed. For non-QS attractive particles, Refs.~\cite{Redner2013Jul,Sarkar2021Feb} identified a reentrant behavior, with a region in their phase diagrams that resembles Region~I. As we demonstrate below, however, QS interactions in dilute regions of the system give rise to effects beyond those observed with attractive interactions.

Because kinetics are slow in Region~I, we test the effect of initial conditions by starting from a compact circular cluster with interface particles pointing inward. Fig.~\ref{fig:phdiag_tau_alpha}e shows that, for low $\alpha$, the phase diagram is unchanged, while at high $\alpha$, a strong QS slowing-down traps particles, producing long-lived clusters. At high $\tau_R$, some particles eventually escape, leading to rearrangements that make clusters less compact, though no homogeneous state appears now. At low $\tau_R$, the cluster persists but develops holes. This is because particles escape and then reattach at random locations, nucleating finger-like protrusions that slowly merge with each other, thus leaving holes that are the precursors of a gel-like structure.

On extremely long timescales~\cite{Omar2021May}, memory of the initial condition ultimately vanishes, with $\fc$ converging to similar final values for both initial preparations (see SM~\cite{SM} for longer simulations, including Movie~S3). At such long times, for high $\tau_R$ and $\alpha$, particles can eventually escape, collide in the gas phase, and assemble into long-lived microclusters. The original cluster then fully dissipates and reorganizes into a nearly homogeneous state. Because these relaxation times are practically inaccessible, we describe the system in terms of two {\it effective} equilibria, a perspective that emerges intrinsically in our theory below, that is, without being imposed. Notably, such effective, initial-condition–dependent phase behavior strongly resembles patterns observed in mussel bed experiments~\cite{vandeKoppel2008Oct,Liu2013Jul}.

\begin{figure}
    \includegraphics[width=\linewidth]{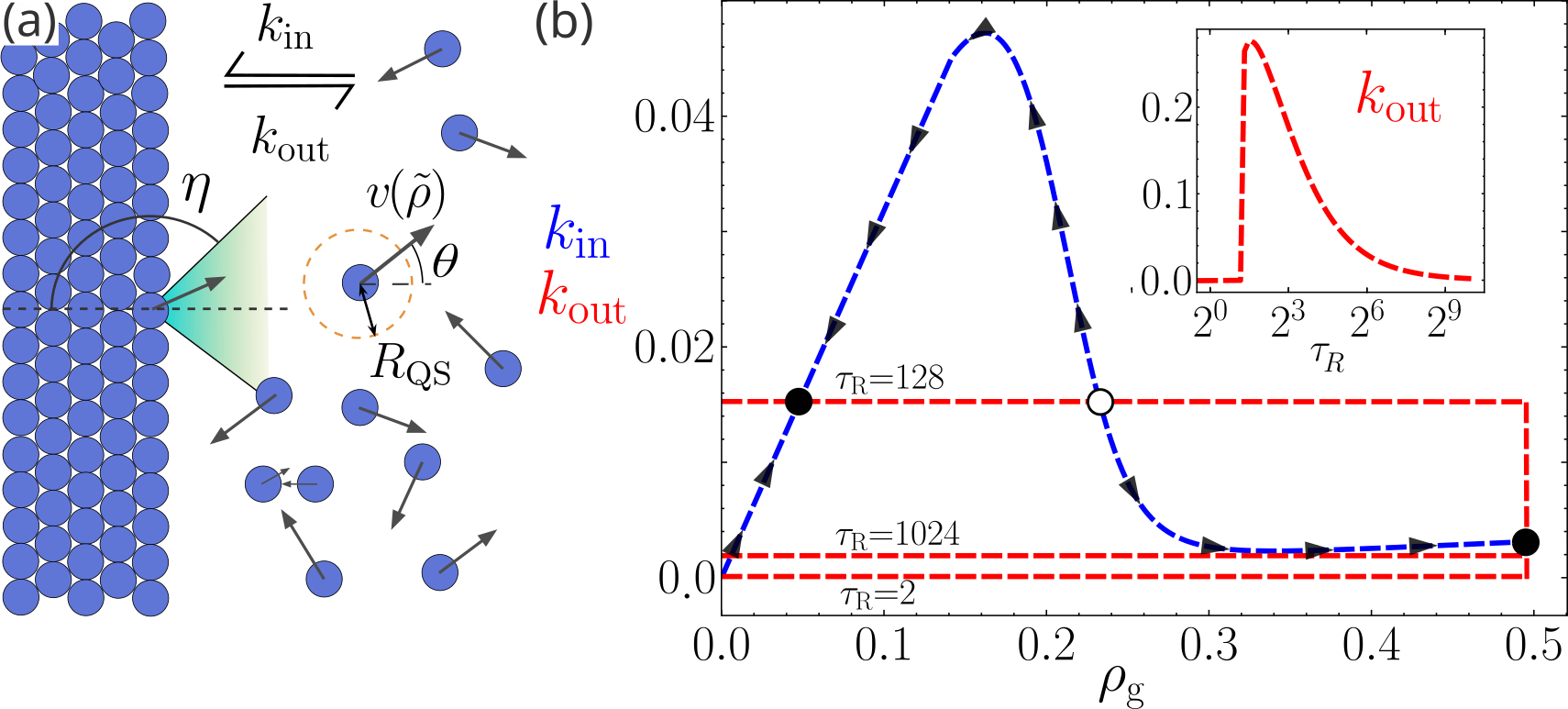}
    \caption{\justifying
    QS effects on the cluster surface and in the gas explain reentrant clustering and initial-condition dependence. 
    (a) Schematics showing effective escape cone: due to an effective attraction, interface particles pointing outside the cone lack sufficient horizontal self-propulsion to escape.
    % $\kin$ and $\kout$ are the absorption and emission rates. 
    % $\eta$ is an auxiliary angle.
    (b) $\kin$ (blue) and $\kout$ (red) vs.\ $\rhog$ [Eqs.~\eqref{eq:kin} and \eqref{eq:kout}] for ${\alpha = 2}$ and $\tauR = 2$, $128$, and $1024$.
    % (b) $\kin$ (blue) and $\kout$ (red) vs.\ $\rhog$ [Eqs.~\eqref{eq:kin} and \eqref{eq:kout}] for ${\alpha = 1}$ and $\tauR = 8$, $128$, and $512$. 
    Inset: $\kout$ vs.\ $\tauR$.
    % Only $\kout$ depends on $\tauR$.
    Quorum sensing makes $\kin$ nontrivial, enabling multiple fixed points $\rhog$. 
    For $\tauR=2$, one has $\kout \approx 0$, giving $\rhog \approx 0$ (Region I clusters). 
    For $\tauR=128$, if ${\rhog \approx \rhoglobal=0.49}$ initially, then $\kout>\kin$, maintaining ${\rhog \approx \rho_0}$ (Region III, homogeneous). 
    For $\tauR=1024$, the system flows to a stable fixed point with low $\rhog$ (Region II clusters). 
    Fitting parameters $C = 1.32$ and $\kappa = 2$ were set by visual comparison of the full phase diagrams.
    % Plots of $\kin$ (blue) and  $\kout$ (red).
    % In specific regions of the phase diagram $\tauR-\alpha$, the stability condition ${\kin=\kout}$ presents up to three solutions $\rhog^*$, where the extreme ones are stable.
    % From left to right, $\rhog^* \approx 0.03, \, 0.24 \text{ and } \rho_{\rm o}$, where $\rho_{\rm o}\approx 0.49$ is the overall density of the system (${\phi=0.4}$).
    % The curves were obtained for $\tauR=256$ and $\alpha=3.5$, with fitting parameters $F_0 = 8.1$ and $\kappa=4.1$.
    % The inset snapshots are from simulations initiated in a homogeneous (crimson, $\fc \approx 0.25$) and heterogeneous (light blue, $\fc \approx 0.75$) configurations. 
    }
    \label{fig:rates}
\end{figure}

{\it Kinetic balance theory}---To properly understand the clustering behavior, we formulate a kinetic theory, following Refs.~\cite{Redner2013Jan,Redner2013Jul}, now for QS interactions. 
A more detailed derivation is in the SM~\cite{SM}. 
The theory calculates the number of particles in each phase in order to derive $\fc$.
We assume a steady state consisting of a single large cluster at (number) density $\rhoc$ coexisting with a homogeneous gas of density $\rhog$; see Fig.~\ref{fig:rates}a. 
We take the theoretical value
$\rhoc = \frac{2}{\sigma^2\sqrt{3}}$, corresponding to close packing in a hexagonal arrangement of touching disks~\cite{Redner2013Jan}.
%, with the cluster sufficiently large to appear planar to particles (Fig.~\ref{fig:rates}a).
Cluster stability is set by the condition $\kout=\kin$, where $\kout$ and $\kin$ denote the particle emission and absorption rates per unit length.

To estimate $\kin$, we compute the flux of gas particles toward the cluster. Following Ref.~\cite{Redner2013Jan}, $\kin$ is proportional to the gas density and speed, except that here the self-propulsion speed depends on $\rhog$ via QS. 
This gives \cite{SM}
\be \label{eq:kin}
\kin = \frac{1}{\pi} \rhog \, v\left( \rhog^{\rm QS} \right),
\ee
using the local approximation ${v[\tilde{\rho}(\mathbf{r})] \simeq v(\rho)}$.
The focal particle perceives a density $\rhog^{\rm QS}=\rhog-1/\AQS$, where its own contribution $1/\AQS$ is excluded.
% The focal particle’s own contribution $1/\AQS$ is discounted from the QS-detected density. 
Collisions are neglected in the gas, but QS is retained due to its longer interaction range. As a result, $\kin$ depends non-monotonically on $\rhog$ (Fig.~\ref{fig:rates}b). Such non-monotonic behavior is absent without QS, where $v=\vmax$~\cite{Redner2013Jan}. QS thus introduces the following feedback: $\rhog$ sets $v$, which in turn modifies $\rhog$ until a balance is reached. As we will see, $\kout$ does not depend on $\rhog$. When $\rhog>\bar{\rho}$, gas particles move slower, whereas for $\rhog<\bar{\rho}$ they move faster (see Fig.~\ref{fig:phdiag_tau_alpha}b); yet, both cases can yield the same $\kin$. This feedback produces up to two stable solutions of $\kin=\kout$, corresponding to high- and low-density gases, and hence to two effective equilibrium values of $\fc$, obtained below.

To derive an approximation for $\kout$, consider a particle at the cluster interface. Once it points outward, the particle starts moving away, but its speed can be very low due to QS. For the particle to escape, we assume it must maintain an outward orientation until it reaches a distance where QS no longer acts. To capture this picture, we take a simple approach. Suppose the particle points at an angle $\theta$. If the direction does not change significantly, the particle escapes. The characteristic time for angular reorientation is $\tauR$. Assuming constant $\theta$, we define $T_{\rm QS}$ as the time required to cross the QS influence zone. The condition for escape is therefore $T_{\rm QS}<\tauR$.  

To estimate $T_{\rm QS}$, we use a simple two-particle scenario, with one particle fixed and the other moving. The speed, denoted $v_{\rm p}(r)$, increases with the interparticle distance $r$ since the weighted detected density decays with $r$. Using the weighting kernel described in the SM \cite{SM} and the QS motility $v(\tilde{\rho})$, we obtain $v_{\rm p}(r)$. In the cluster-particle scenario, only the $x$-axis velocity component matters (Fig.~\ref{fig:rates}a), so we take $v_{\rm p}(x)\cos\theta$. We also multiply by a universal fitting constant $C$ to partially account for our simplified approach. The crossing time is then
\begin{equation}
T_{\rm QS}=\int_{\sigma}^{\RQS}\frac{dx}{C\,v_{\rm p}(x)\cos\theta}.
\end{equation}
The integral runs from the initial separation, $x=\sigma$, up to $x=\RQS$, beyond which QS effects vanish.

Below a threshold value of $\theta$, escape occurs. At the threshold, one has
\be \label{eq:escape_cone}
\tauR \cos(\pi-\eta) = \int_{\sigma}^{\RQS}\frac{dx}{C\,v_{\rm p}(x)},
\ee
where ${\eta \in (\frac{\pi}{2},\pi)}$ defines a ``escape cone'' with half aperture $\pi-\eta$ (Fig.~\ref{fig:rates}a).
%A particle at the cluster interface $\theta<$ orientation will escape.
We obtain $\eta$ numerically from Eq.~\eqref{eq:escape_cone}. 

To proceed with our calculation of $\kout$, we follow Ref.~\cite{Redner2013Jul} by considering the angular probability distribution of particles on the cluster surface and solving a rotational diffusion problem with absorbing boundaries at the edges of the escape cone defined by $\eta$. This leads to~\cite{SM}
\be \label{eq:kout}
k_\text{out} = \frac{\pi^2 \kappa}{4 \sigma \tauR \eta^2}.
\ee
Here, $\kappa$ denotes the effective number of particles released per escape event: when one particle escapes, it may trigger an avalanche of further escapes. 
For simplicity, and following Refs.~\cite{Redner2013Jan,Redner2013Jul}, $\kappa$ is treated as a universal fitting parameter. 
We note that, without QS, low $\tauR$ leads to high $\kout$, whereas with QS, the behavior depends on the functional dependence of $\eta$ on $\tauR$.
 %In fact, we will see that low $\tauR$ can also produce low $\kout$.
% , and the particles' attraction forms a cluster.
% All these processes lead to small $\kout$ values and can be expressed by
% \be \label{eq:kout}
% k_\text{out} = \frac{\pi^2 \kappa}{4 \sigma \tauR \eta^2}.
% \ee
% A detailed derivation can be found in the Supplementary Material~\cite{SM}.

% We must still express $\UQS$ (or $\eta$) in terms of the QS parameters. To this end, we adopt a standard mean-field approach to MIPS~\cite{Tailleur2008May,Cates2013Feb,Cates2015Mar}, where an effective free energy is defined in terms of a density-dependent self-propulsion speed $v(\rho)$, here caused by QS:
% $ %\label{eq:eff_free_energy_MIPS}
% {f(\rho) = \frac{1}{2} \int \ln \left[ \tauR v^2(s) + 2 \DT \right] \, ds + \rho(\ln\rho-1).}
% $
% The first term arises from microscopic interactions, while the second reflects entropic contributions. By comparison with the mean-field term for attractive particles, $-\tfrac{\UQS}{2}\rho^2$, we obtain
% $
% {\UQS = \frac{ \vmax (\vmax-\vmin) \tauR }{2 \DT + \vmax^2 \tauR} \sech{2}{ \frac{\bar{\rho}}{\delta \rho} } \frac{\alpha}{ \delta \rho}}$.
% For details, see Supplementary Material~\cite{SM}.
% $$
% \eta = \pi - \cos^{-1} \left( F_0 \frac{1}{\vmax \tauR} \frac{ \vmax (\vmax-\vmin) \tauR }{2 \DT + \vmax^2 \tauR} \sech{2}{ \frac{\bar{\rho}}{\delta \rho} } \frac{\alpha}{ \delta \rho} \right)
% $$
% % $$
% % \eta = \pi - \cos^{-1} \left( F_0 \frac{1}{\vmax \tauR} \frac{ \vmax^2 \tauR }{ \vmax^2 \tauR} \sech{2}{ \frac{\bar{\rho}}{\delta \rho} } \frac{\alpha}{ \delta \rho} \right)
% % $$

\begin{figure}
    \centering
     \includegraphics[width=\linewidth]{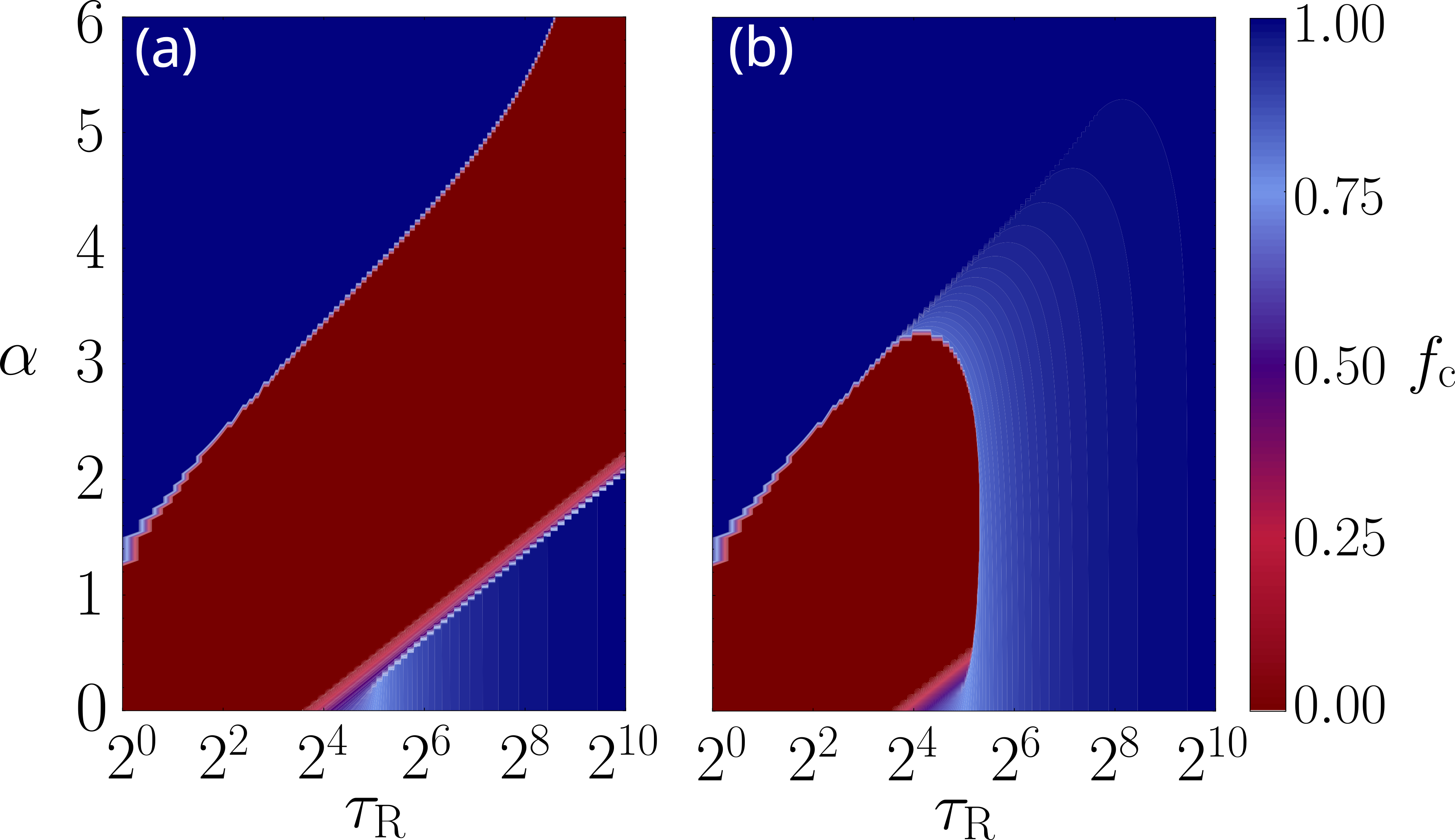}
    \caption{\justifying
    Effective kinetic balance theory captures key features of the clustering phase diagram. 
    Fraction of particles in clusters $\fc$ from our theory [Eq.~\eqref{eq:fc_eff}] is shown for (a) homogeneous and (b) clustered initial conditions. 
    The theory reproduces the reentrant clustering and the initial-condition dependence of the diagram, while finer features such as cluster morphology and the pronounced curvature of the transition lines are not resolved.    % \new{The theory neglects fluctuations, which could lead to the same phase diagram in panels (a) and (b); see Supplementary Material~\cite{SM}.}
    % $\fc$ obtained a, c~from simulations and b, d~from Eq.~\eqref{eq:fc_eff}, as a function of $\tauR$ and $\alpha$.
    % a-b and c-d consist of the homogeneous and heterogeneous initial conditions, respectively.
    % The ``$\times$'' marks represent the simulation points used to interpolate the plot.
    % The fitting parameters are chosen by eye as $F_0 = 8.1$ and $\kappa=4.1$.
    % In the homogeneous initial condition case, the reentrant behaviors by either tuning $\tauR$ or $\alpha$ are captured by Eq.~\eqref{eq:fc_eff}.
    % The region where phase separation occurs is also observed in the heterogeneous initial condition case, but only the reentrant behavior on $\tauR$ is captured.
    }
    \label{fig:fc_eff}
\end{figure}

Using $\kin=\kout$ and conservation of the total number of particles, we can express the clustering order parameter:
\be \label{eq:fc_eff}
\fc = \frac{16 \phi \eta^2 v(\rhog^{\rm QS}) \tauR - \pi^4 \sigma \kappa}{16 \phi \eta^2 v(\rhog^{\rm QS}) \tauR - 2\sqrt{3} \pi^3 \phi \sigma \kappa}.
\ee
% \rhog^{\rm QS} = \rhog-1/\AQS
% with
% \be
% \eta = \pi - \cos^{-1} \left( \frac{F_0 \, (\vmax-\vmin) }{2 D_{\rm T} + v_{\rm max}^2 \tauR } \operatorname{sech}^2 \left( \frac{\bar{\rho}}{\delta\rho} \right) \frac{\alpha}{\delta\rho} \right).
% \ee
%
%
% As we show in Fig.~\ref{fig:rates}b, due to $v(\rhog)$ in Eqs.~\eqref{eq:kin}, the stability condition leads to at most two stable solutions $\rhog^*$, representing the two effective equilibria found in Fig.~\ref{fig:phdiag_tau_alpha}d and e.
%
This is the same as the $\fc$ expression in Ref.~\cite{Redner2013Jul}, except that now $v(\rhog^{\rm QS})$ and $\eta$ are affected by QS. Figure~\ref{fig:fc_eff} shows the result. Despite approximations, the theory reproduces both the reentrant and the initial-condition–dependent behaviors. To illustrate how these features arise in the theory, consider $\alpha=2$ in Fig.~\ref{fig:fc_eff}a and notice $\kin(\rhog)$ in Fig.~\ref{fig:rates}b. For low $\tauR$, particles are too non-persistent and too slow to escape QS, corresponding to a fully closed escape cone ($\eta=\pi$) and $\kout\approx0$ (inset of Fig.~\ref{fig:rates}b). This yields $\rhog\approx0$ and $\fc=1$. As $\tauR$ crosses a first critical value, $\kout$ rises sharply (since now $\pi/2<\eta<\pi$) and the gas will have the global density $\rhog\approx\rhoglobal$, giving $\fc\approx0$. For sufficient $\tauR$, two solutions emerge. If the system starts near $\rhog\approx\rhoglobal$, then $\kout>\kin$ keeps it at the nearby solution, $\rhog\approx\rhoglobal$: persistence now enables particles to escape QS. If instead the system starts near $\rhog\approx0$, $\kin$ grows with $\rhog$ until a low-$\rhog$ balance is reached, yielding another stable solution. 
In the simulations, fluctuations not captured by the theory progressively steer regions of the system toward the first solution. 
Upon further increasing $\tauR$ beyond a second critical value, contact-force MIPS lowers $\kout$ again, restoring $\rhog\approx0$ and $\fc\approx1$, despite the wide open escape cone ($\eta\approx\pi/2$).

\begin{figure}
    \centering
    \includegraphics[width=\linewidth]{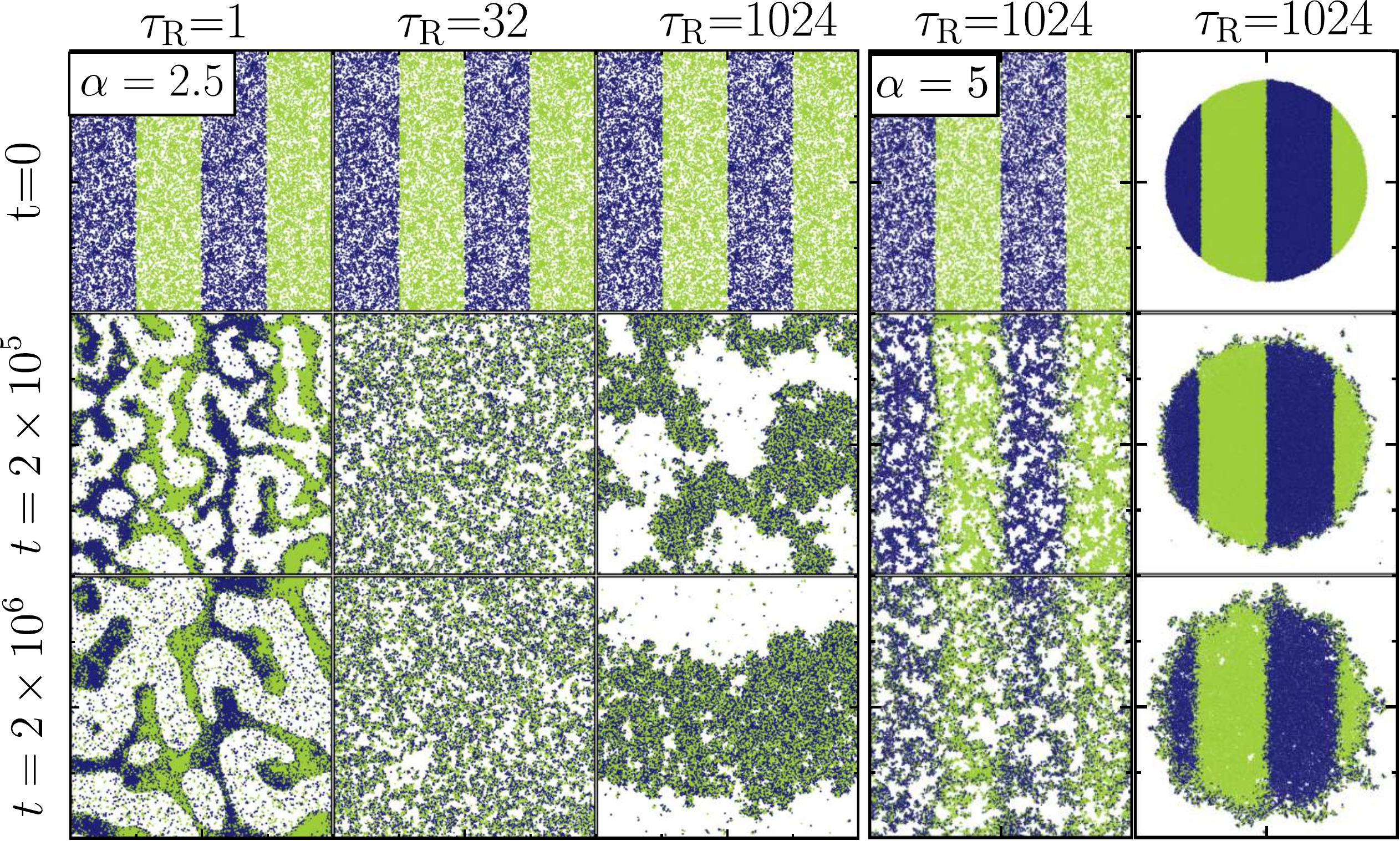}
    \caption{\justifying
    % Phase separation evolution for multiple a~$\tauR$ and b~$\alpha$.
    % The colored stripes virtually represent the particles' initial positions.
    % In [a~bottom row]  [b~top row] emerges a kinetic arrest, represented by the lack of mixing.
    % As $\tauR$ is decreased in a (or $\alpha$ increased in b), the system goes through a reentrant clustering transition characterized by kinetic arrest dynamics.
    % Note that the colored stripes are preserved for $\tauR=1$ (or $\alpha = 5$), indicating that particles were almost static during system evolution since their motion is limited by chemical attraction and mechanical repulsion.
    % In b, the dynamics of $\alpha=0 \text{ and } 1$ are distinguished by the reduction of the dilute phase and the fluctuation in the cluster size and shape; in $\alpha=1$, the cluster can split and merge, constantly changing its shape.
    Quorum-sensing kinetic arrest manifests as suppressed mixing and strong memory of initial conditions.
    Particles are colored by their initial positions, grouped in vertical stripes with alternating colors. The first, second, and third columns show time evolution for $\alpha = 2.5$ and $\tauR = 1$, $32$, and $1024$, respectively. The fourth and fifth columns correspond to $\alpha = 5$ and $\tauR = 1024$, starting from homogeneous and clustered initial conditions. Preservation of colored stripes over time indicates minimal rearrangement. Mixing slows down with increasing $\alpha$, decreasing $\tauR$, or stronger initial clustering.
    }
    \label{fig:snap_mixing}
\end{figure}

{\it Kinetic arrest}---The physics above stems from the strong motility reduction induced by QS, which produces kinetically arrested configurations. To probe this arrest, we examine the suppression of particle mixing, reflected in a strong memory of initial conditions.
Particles are colored by their initial positions in alternating stripes and the system is then evolved (Fig.~\ref{fig:snap_mixing}). When QS dominates ($\alpha=2.5$, $\tauR=1$), particles accumulate near their neighbors, forming nearly static microclusters that slowly deform into a gel. The persistence of the initial stripes over long timescales confirms the limited particle motion.

As $\tauR$ increases, persistence allows particles to escape QS, causing gels to lose structure. The resulting motility destroys kinetic arrest, as shown by the enhanced mixing for $\alpha=2.5$ at $\tauR=32$ and $1024$ (Fig.~\ref{fig:snap_mixing}). By further increasing QS to $\alpha=5$ with $\tauR=1024$, the dynamics exhibit a renewed slowdown. For homogeneous initial states, stripes fade more quickly than for single-cluster initial conditions, where caging delays escape. %showing the influence of initial conditions on the lifetime of these kinetically arrested states.

%%%%%%%%%%%%%%%%%%%%%%%%%%%%%%%%%

{\it Conclusions}---We have shown how quorum sensing shapes clustering in active matter. The formation of quorum-sensing active gels, which are destroyed when self-propulsion is highly persistent, gives rise to a reentrant clustering phase diagram. Depending on the initial condition, kinetically arrested configurations lead to distinct effective equilibria. A kinetic balance theory captures both the reentrant behavior and the origin of the two equilibria. Our results highlight parallels between synthetic and biological systems and establish quorum sensing as a key ingredient in active matter self-organization.
%We concluded that these effective equilibria are a direct consequence of kinetically arrested configurations induced by QS and contact forces.
%Unlike active systems with passive attraction, where gels are a transient configuration and a single cluster is the final state~\cite{Redner2013Jul,Sarkar2021Feb}, with QS, gels are the most stable configuration.
%
%Indeed, this class of system represents a novel active matter framework, little explored so far~\cite{Jose2021Mar,Thapa2024Feb,Lefranc2025Aug,Nguyen2025Jul}.

Several future directions arise, including the study of non-reciprocal QS interactions~\cite{Agudo-Canalejo2019Jul,Dinelli2023Nov}, the development of a hydrodynamic theory analysis combining QS and steric effects~\cite{Lefranc2025Aug}, and the exploration of ecological implications of clustering~\cite{Forget2022Dec,zusman2007chemosensory}. In addition, models that explicitly include signaling molecules may uncover novel crowding effects. Ultimately, one should compare our results with experiments in bacterial systems~\cite{Curatolo2020Nov}.

We thank the Brazilian agencies CAPES, CNPq, and FAPESP. 
%H.C.M.F.\ and L.G.B.\ acknowledge CNPq (procs.\ 402487/2023-0 and 443517/2023-1).
L.d.S.\ was supported by Scholarships No.\ 140921/2021-4 (CNPq) and No.\ 88887.911602/2023-00 (CAPES).
E.F.T.\ acknowledges ICTP-SAIFR/IFT-UNESP.
%The simulations used the \href{https://pnipe.mcti.gov.br/laboratory/19775}{VD Lab} cluster at IF-UFRGS. 
P.d.C.\ was supported by Scholarships No.\ 2021/10139-2 and No.\ 2022/13872-5 and ICTP-SAIFR Grant No.\ 2021/14335-0, all granted by São Paulo Research Foundation (FAPESP), Brazil.
% To some extent, QS can be seen as a mimic of a chemically interacting biological system.
% Hence, this work represents a step toward better understanding systems such as bacteria~\cite{Liu2019Jun,Curatolo2020Nov,vanGestel2021Apr} and ants~\cite{Anderson2022Nov} colonies and mussel beds~\cite{Liu2013Jul}.
% Particularly, the observed effective equilibria present a resemblance to experiments with the latter.
% It can also serve as a framework to understand how bio-inspired synthetic systems~\cite{Wang2024Oct,Lefranc2025Aug} can be developed.
%

%%%%%%%%%%%%%%%%%%%%%%%%%%%%%%%%%

% \begin{acknowledgements}
%     Blabla
% \end{acknowledgements}

% \nocite{*}
% \clearpage
\bibliographystyle{apsrev4-2}
\bibliography{QS_EV}% Produces the bibliography via BibTeX.

@article{PhysRevLett.134.138401,
  title = {Segregation in Binary Mixture with Differential Contraction among Active Rings},
  author = {Teixeira, Emanuel F. and Beatrici, Carine P. and Fernandes, Heitor C. M. and Brunnet, Leonardo G.},
  journal = {Phys. Rev. Lett.},
  volume = {134},
  issue = {13},
  pages = {138401},
  numpages = {6},
  year = {2025},
  month = {Mar},
  publisher = {American Physical Society},
  doi = {10.1103/PhysRevLett.134.138401},
  url = {https://link.aps.org/doi/10.1103/PhysRevLett.134.138401}
}

@article{Dinelli2023Nov,
	author = {Dinelli, Alberto and O{'}Byrne, J{\ifmmode\acute{e}\else\'{e}\fi}r{\ifmmode\acute{e}\else\'{e}\fi}my and Curatolo, Agnese and Zhao, Yongfeng and Sollich, Peter and Tailleur, Julien},
	title = {{Non-reciprocity across scales in active mixtures}},
	journal = {Nat. Commun.},
	volume = {14},
	number = {7035},
	pages = {1--10},
	year = {2023},
	month = {nov},
	issn = {2041-1723},
	publisher = {Nature Publishing Group},
	doi = {10.1038/s41467-023-42713-5}
}

@article{Henkes2011Oct,
	author = {Henkes, Silke and Fily, Yaouen and Marchetti, M. Cristina},
	title = {{Active jamming: Self-propelled soft particles at high density}},
	journal = {Phys. Rev. E},
	volume = {84},
	number = {4},
	pages = {040301},
	year = {2011},
	month = oct,
	issn = {2470-0053},
	publisher = {American Physical Society},
	doi = {10.1103/PhysRevE.84.040301}
}

@article{Fily2012Jun,
	author = {Fily, Yaouen and Marchetti, M. Cristina},
	title = {{Athermal Phase Separation of Self-Propelled Particles with No Alignment}},
	journal = {Phys. Rev. Lett.},
	volume = {108},
	number = {23},
	pages = {235702},
	year = {2012},
	month = jun,
	issn = {1079-7114},
	publisher = {American Physical Society},
	doi = {10.1103/PhysRevLett.108.235702}
}

@article{rojas2023wetting,
	author = {Rojas-Vega, Mauricio and de Castro, Pablo and Soto, Rodrigo},
	title = {{Mixtures of self-propelled particles interacting with asymmetric obstacles}},
	journal = {Eur. Phys. J. E},
	volume = {46},
	number = {10},
	pages = {1--11},
	year = {2023},
	month = oct,
	issn = {1292-895X},
	publisher = {Springer Berlin Heidelberg},
	doi = {10.1140/epje/s10189-023-00354-y}
}

@article{teixeira2025collective,
  title         = {Collective ballistic motion explains fast aggregation in adhesive active matter},
  author        = {Teixeira, Emanuel F. and de Castro, Pablo and Beatrici, Carine P. and Fernandes, Heitor C. M. and Brunnet, Leonardo G.},
  year          = {2025},
  url           = {https://arxiv.org/pdf/2508.11793},
  eprint        = {2508.11793},
  journal       = {arXiv},
  primaryclass  = {astro-ph.IM}
}

@article{villa2020run,
  title={Run-and-tumble bacteria slowly approaching the diffusive regime},
  author={Villa-Torrealba, Andrea and Ch{\'a}vez-Raby, Crist{\'o}bal and de Castro, Pablo and Soto, Rodrigo},
  journal={Physical Review E},
  volume={101},
  number={6},
  pages={062607},
  year={2020},
  publisher={APS}
}

@book{van2024soft,
  title     = {Soft Matter: Concepts, Phenomena, and Applications},
  author    = {Wim van Saarloos and Vincenzo Vitelli and Zorana Zeravcic},
  publisher = {Princeton University Press},
  year      = {2024},
  address   = {Princeton, NJ},
  isbn      = {9780691191300},
  url       = {https://press.princeton.edu/books/hardcover/9780691191300/soft-matter}
}

@article{zusman2007chemosensory,
  title={Chemosensory pathways, motility and development in Myxococcus xanthus},
  author={Zusman, David R and Scott, Ansley E and Yang, Zhaomin and Kirby, John R},
  journal={Nature Reviews Microbiology},
  volume={5},
  number={11},
  pages={862--872},
  year={2007},
  publisher={Nature Publishing Group UK London}
}

@article{daniels2004quorum,
  title         = {Quorum sensing and swarming migration in bacteria},
  author        = {Daniels, Ruth and Vanderleyden, Jos and Michiels, Jan},
  year          = {2004},
  month         = jun,
  journal       = {FEMS Microbiology Reviews},
  publisher     = {Oxford University Press (OUP)},
  volume        = {28},
  number        = {3},
  pages         = {261–289},
  doi           = {10.1016/j.femsre.2003.09.004},
  issn          = {1574-6976},
  url           = {http://dx.doi.org/10.1016/j.femsre.2003.09.004}
}

@article{ridgway2023motility,
  title         = {Motility-Induced Phase Separation Mediated by Bacterial Quorum Sensing},
  author        = {Ridgway, W. J. M. and Dalwadi, M. P. and Pearce, P. and Chapman, S. J.},
  year          = {2023},
  month         = nov,
  journal       = {Physical Review Letters},
  publisher     = {American Physical Society (APS)},
  volume        = {131},
  number        = {22},
  issn          = {1079-7114},
  url           = {http://dx.doi.org/10.1103/PhysRevLett.131.228302}
}

@article{palacci2013living,
  title         = {Living Crystals of Light-Activated Colloidal Surfers},
  author        = {Palacci, J. and Sacanna, S. and Steinberg, A. P. and Pine, D. J. and Chaikin, P. M.},
  year          = {2013},
  month         = feb,
  journal       = {Science},
  publisher     = {American Association for the Advancement of Science (AAAS)},
  volume        = {339},
  number        = {6122},
  pages         = {936–940},
  doi           = {10.1126/science.1230020},
  issn          = {1095-9203},
  url           = {http://dx.doi.org/10.1126/science.1230020}
}

@article{bechinger2016active,
	author = {Bechinger, Clemens and Di Leonardo, Roberto and L{\ifmmode\ddot{o}\else\"{o}\fi}wen, Hartmut and Reichhardt, Charles and Volpe, Giorgio and Volpe, Giovanni},
	title = {{Active particles in complex and crowded environments}},
	journal = {Rev. Mod. Phys.},
	volume = {88},
	number = {4},
	pages = {045006},
	year = {2016},
	month = nov,
	issn = {1539-0756},
	publisher = {American Physical Society},
	doi = {10.1103/RevModPhys.88.045006}
}

@article{Redner2013Jan,
	author = {Redner, G. S. and Hagan, M. F. and Baskaran, A.},
	title = {{Structure and Dynamics of a Phase-Separating Active Colloidal Fluid}},
	journal = {Phys. Rev. Lett.},
	volume = {110},
	number = {5},
	pages = {055701},
	year = {2013},
	month = jan,
	issn = {1079-7114},
	publisher = {American Physical Society},
	doi = {10.1103/PhysRevLett.110.055701}
}

@article{Sarkar2021Feb,
	author = {Sarkar, D. and Gompper, G. and Elgeti, J.},
	title = {{A minimal model for structure, dynamics, and tension of monolayered cell colonies}},
	journal = {Commun. Phys.},
	volume = {4},
	number = {36},
	pages = {1--8},
	year = {2021},
	month = feb,
	issn = {2399-3650},
	publisher = {Nature Publishing Group},
	doi = {10.1038/s42005-020-00515-x}
}

@article{Redner2013Jul,
	author = {Redner, G. S. and Baskaran, A. and Hagan, M. F.},
	title = {{Reentrant phase behavior in active colloids with attraction}},
	journal = {Phys. Rev. E},
	volume = {88},
	number = {1},
	pages = {012305},
	year = {2013},
	month = jul,
	issn = {2470-0053},
	publisher = {American Physical Society},
	doi = {10.1103/PhysRevE.88.012305}
}

@article{Jose2021Mar,
	author = {Jose, F. and Anand, S. K. and Singh, S. P.},
	title = {{Phase separation of an active colloidal suspension via quorum-sensing}},
	journal = {Soft Matter},
	volume = {17},
	number = {11},
	pages = {3153--3161},
	year = {2021},
	month = mar,
	issn = {1744-683X},
	publisher = {The Royal Society of Chemistry},
	doi = {10.1039/D0SM02131H}
}

@article{vanGestel2021Apr,
	author = {van Gestel, J. and Bareia, T. and Tenennbaum, B. and Dal Co, A. and Guler, P. and Aframian, N. and Puyesky, S. and Grinberg, I. and D{'}Souza, G. G. and Erez, Z. and Ackermann, M. and Eldar, A.},
	title = {{Short-range quorum sensing controls horizontal gene transfer at micron scale in bacterial communities}},
	journal = {Nat. Commun.},
	volume = {12},
	number = {2324},
	pages = {1--11},
	year = {2021},
	month = apr,
	issn = {2041-1723},
	publisher = {Nature Publishing Group},
	doi = {10.1038/s41467-021-22649-4}
}

@article{DalCo2020Mar,
	author = {Dal Co, A. and van Vliet, S. and Kiviet, D. J. and Schlegel, S. and Ackermann, M.},
	title = {{Short-range interactions govern the dynamics and functions of microbial communities}},
	journal = {Nat. Ecol. Evol.},
	volume = {4},
	number = {3},
	pages = {366--375},
	year = {2020},
	month = mar,
	issn = {2397-334X},
	publisher = {Nature Publishing Group},
	doi = {10.1038/s41559-019-1080-2}
}

@article{Cates2015Mar,
	author = {Cates, M. E. and Tailleur, J.},
	title = {{Motility-Induced Phase Separation}},
	journal = {Annu. Rev. Condens. Matter Phys.},
	number = {Volume 6, 2015},
	pages = {219--244},
	year = {2015},
	month = mar,
	publisher = {Annual Reviews},
	doi = {10.1146/annurev-conmatphys-031214-014710}
}

@article{Tong2019Dec,
	author = {Tong, H. and Tanaka, H.},
	title = {{Structural order as a genuine control parameter of dynamics in simple glass formers}},
	journal = {Nat. Commun.},
	volume = {10},
	number = {5596},
	pages = {1--10},
	year = {2019},
	month = dec,
	issn = {2041-1723},
	publisher = {Nature Publishing Group},
	doi = {10.1038/s41467-019-13606-3}
}

@article{Sokolov2007Apr,
	author = {Sokolov, A. and Aranson, I. S. and Kessler, J. O. and Goldstein, R. E.},
	title = {{Concentration Dependence of the Collective Dynamics of Swimming Bacteria}},
	journal = {Phys. Rev. Lett.},
	volume = {98},
	number = {15},
	pages = {158102},
	year = {2007},
	month = apr,
	publisher = {American Physical Society},
	doi = {10.1103/PhysRevLett.98.158102}
}

@article{Ballerini2008Jan,
	author = {Ballerini, M. and Cabibbo, N. and Candelier, R. and Cavagna, A. and Cisbani, E. and Giardina, I. and Lecomte, V. and Orlandi, A. and Parisi, G. and Procaccini, A. and Viale, M. and Zdravkovic, V.},
	title = {{Interaction ruling animal collective behavior depends on topological rather than metric distance: Evidence from a field study}},
	journal = {Proc. Natl. Acad. Sci. U.S.A.},
	volume = {105},
	number = {4},
	pages = {1232--1237},
	year = {2008},
	month = jan,
	publisher = {Proceedings of the National Academy of Sciences},
	doi = {10.1073/pnas.0711437105}
}

@article{Buttinoni2013Jun,
	author = {Buttinoni, I. and Bialk{\ifmmode\acute{e}\else\'{e}\fi}, J. and K{\ifmmode\ddot{u}\else\"{u}\fi}mmel, F. and L{\ifmmode\ddot{o}\else\"{o}\fi}wen, H. and Bechinger, C. and Speck, T.},
	title = {{Dynamical Clustering and Phase Separation in Suspensions of Self-Propelled Colloidal Particles}},
	journal = {Phys. Rev. Lett.},
	volume = {110},
	number = {23},
	pages = {238301},
	year = {2013},
	month = jun,
	publisher = {American Physical Society},
	doi = {10.1103/PhysRevLett.110.238301}
}

@article{Fischer2020Jan,
	author = {Fischer, A. and Schmid, F. and Speck, T.},
	title = {{Quorum-sensing active particles with discontinuous motility}},
	journal = {Phys. Rev. E},
	volume = {101},
	number = {1},
	pages = {012601},
	year = {2020},
	month = jan,
	publisher = {American Physical Society},
	doi = {10.1103/PhysRevE.101.012601}
}

@article{Anderson2022Nov,
	author = {Anderson, C. and Fernandez-Nieves, A.},
	title = {{Social interactions lead to motility-induced phase separation in fire ants}},
	journal = {Nat. Commun.},
	volume = {13},
	number = {6710},
	pages = {1--10},
	year = {2022},
	month = nov,
	issn = {2041-1723},
	publisher = {Nature Publishing Group},
	doi = {10.1038/s41467-022-34181-0}
}

@article{Gordon1993Jun,
	author = {Gordon, D. M. and Paul, R. E. and Thorpe, K.},
	title = {{What is the function of encounter patterns in ant colonies?}},
	journal = {Anim. Behav.},
	volume = {45},
	number = {6},
	pages = {1083--1100},
	year = {1993},
	month = jun,
	issn = {0003-3472},
	publisher = {Academic Press},
	doi = {10.1006/anbe.1993.1134}
}

@article{Duan2023Oct,
	author = {Duan, Y. and Agudo-Canalejo, J. and Golestanian, R. and Mahault, B.},
	title = {{Dynamical Pattern Formation without Self-Attraction in Quorum-Sensing Active Matter: The Interplay between Nonreciprocity and Motility}},
	journal = {Phys. Rev. Lett.},
	volume = {131},
	number = {14},
	pages = {148301},
	year = {2023},
	month = oct,
	publisher = {American Physical Society},
	doi = {10.1103/PhysRevLett.131.148301}
}

@article{Forget2022Dec,
	author = {Forget, Mathieu and Adiba, Sandrine and Brunnet, Leonardo Gregory and De Monte, Silvia},
	title = {{Heterogeneous individual motility biases group composition in a model of aggregating cells}},
	journal = {Front. Ecol. Evol.},
	volume = {10},
	pages = {1052309},
	year = {2022},
	month = dec,
	issn = {2296-701X},
	publisher = {Frontiers},
	doi = {10.3389/fevo.2022.1052309}
}

@article{Bauerle2018Aug,
	author = {B{\ifmmode\ddot{a}\else\"{a}\fi}uerle, Tobias and Fischer, Andreas and Speck, Thomas and Bechinger, Clemens},
	title = {{Self-organization of active particles by quorum sensing rules}},
	journal = {Nat. Commun.},
	volume = {9},
	number = {3232},
	pages = {1--8},
	year = {2018},
	month = aug,
	issn = {2041-1723},
	publisher = {Nature Publishing Group},
	doi = {10.1038/s41467-018-05675-7}
}

@article{Agudo-Canalejo2019Jul,
	author = {Agudo-Canalejo, Jaime and Golestanian, Ramin},
	title = {{Active Phase Separation in Mixtures of Chemically Interacting Particles}},
	journal = {Phys. Rev. Lett.},
	volume = {123},
	number = {1},
	pages = {018101},
	year = {2019},
	month = jul,
	issn = {1079-7114},
	publisher = {American Physical Society},
	doi = {10.1103/PhysRevLett.123.018101}
}

@article{Curatolo2020Nov,
	author = {Curatolo, A. I. and Zhou, N. and Zhao, Y. and Liu, C. and Daerr, A. and Tailleur, J. and Huang, J.},
	title = {{Cooperative pattern formation in multi-component bacterial systems through reciprocal motility regulation}},
	journal = {Nat. Phys.},
	volume = {16},
	pages = {1152--1157},
	year = {2020},
	month = nov,
	issn = {1745-2481},
	publisher = {Nature Publishing Group},
	doi = {10.1038/s41567-020-0964-z}
}

@article{Liu2013Jul,
	author = {Liu, Quan-Xing and Doelman, Arjen and Rottsch{\ifmmode\ddot{a}\else\"{a}\fi}fer, Vivi and de Jager, Monique and Herman, Peter M. J. and Rietkerk, Max and van de Koppel, Johan},
	title = {{Phase separation explains a new class of self-organized spatial patterns in ecological systems}},
	journal = {Proc. Natl. Acad. Sci. U.S.A.},
	volume = {110},
	number = {29},
	pages = {11905--11910},
	year = {2013},
	month = jul,
	publisher = {Proceedings of the National Academy of Sciences},
	doi = {10.1073/pnas.1222339110}
}

@article{vandeKoppel2008Oct,
	author = {van de Koppel, J. and Gascoigne, J. C. and Theraulaz, G. and Rietkerk, M. and Mooij, W. M. and Herman, P. M. J.},
	title = {{Experimental Evidence for Spatial Self-Organization and Its Emergent Effects in Mussel Bed Ecosystems}},
	journal = {Science},
	volume = {322},
	number = {5902},
	pages = {739--742},
	year = {2008},
	month = oct,
	issn = {0036-8075},
	publisher = {American Association for the Advancement of Science},
	doi = {10.1126/science.1163952}
}

@article{Lefranc2025Aug,
  title         = {Synthetic Quorum Sensing and Absorbing Phase Transitions in Colloidal Active Matter},
  author        = {Lefranc, T. and Dinelli, A. and Fernández-Rico, C. and Dullens, R. P. A. and Tailleur, J. and Bartolo, D.},
  year          = {2025},
  month         = aug,
  journal       = {Physical Review X},
  publisher     = {American Physical Society (APS)},
  volume        = {15},
  number        = {3},
  issn          = {2160-3308},
  url           = {http://dx.doi.org/10.1103/8csn-71jk}
}

@article{Tailleur2008May,
	author = {Tailleur, J. and Cates, M. E.},
	title = {{Statistical Mechanics of Interacting Run-and-Tumble Bacteria}},
	journal = {Phys. Rev. Lett.},
	volume = {100},
	number = {21},
	pages = {218103},
	year = {2008},
	month = may,
	publisher = {American Physical Society},
	doi = {10.1103/PhysRevLett.100.218103}
}

@article{Omar2021May,
	author = {Omar, A. K. and Klymko, K. and GrandPre, T. and Geissler, P. L.},
	title = {{Phase Diagram of Active Brownian Spheres: Crystallization and the Metastability of Motility-Induced Phase Separation}},
	journal = {Phys. Rev. Lett.},
	volume = {126},
	number = {18},
	pages = {188002},
	year = {2021},
	month = may,
	publisher = {American Physical Society},
	doi = {10.1103/PhysRevLett.126.188002}
}

@article{Segre2001Jun,
	author = {Segr{\ifmmode\grave{e}\else\`{e}\fi}, P. N. and Prasad, V. and Schofield, A. B. and Weitz, D. A.},
	title = {{Glasslike Kinetic Arrest at the Colloidal-Gelation Transition}},
	journal = {Phys. Rev. Lett.},
	volume = {86},
	number = {26},
	pages = {6042--6045},
	year = {2001},
	month = jun,
	publisher = {American Physical Society},
	doi = {10.1103/PhysRevLett.86.6042}
}

@article{Merrigan2020Mar,
	author = {Merrigan, C. and Ramola, K. and Chatterjee, R. and Segall, N. and Shokef, Y. and Chakraborty, B.},
	title = {{Arrested states in persistent active matter: Gelation without attraction}},
	journal = {Phys. Rev. Res.},
	volume = {2},
	number = {1},
	pages = {013260},
	year = {2020},
	month = mar,
	publisher = {American Physical Society},
	doi = {10.1103/PhysRevResearch.2.013260}
}

@article{Liu2019Jun,
	author = {Liu, G. and Patch, A. and Bahar, F. and Yllanes, D. and Welch, R. D. and Marchetti, M. C. and Thutupalli, S. and Shaevitz, J. W.},
	title = {{Self-Driven Phase Transitions Drive Myxococcus xanthus Fruiting Body Formation}},
	journal = {Phys. Rev. Lett.},
	volume = {122},
	number = {24},
	pages = {248102},
	year = {2019},
	month = jun,
	publisher = {American Physical Society},
	doi = {10.1103/PhysRevLett.122.248102}
}

@article{Nguyen2025Jul,
	author = {Nguyen, Q. M. and Dinelli, A. and Spera, G. and Tailleur, J.},
	title = {{Contact Forces in Motility-Regulated Active Matter}},
	journal = {arXiv},
	year = {2025},
	month = jul,
	eprint = {2507.08964},
	doi = {10.48550/arXiv.2507.08964}
}

@article{Zhao2023Sep,
	author = {Zhao, H. and Ko{\ifmmode\check{s}\else\v{s}\fi}mrlj, A. and Datta, S. S.},
	title = {{Chemotactic Motility-Induced Phase Separation}},
	journal = {Phys. Rev. Lett.},
	volume = {131},
	number = {11},
	pages = {118301},
	year = {2023},
	month = sep,
	publisher = {American Physical Society},
	doi = {10.1103/PhysRevLett.131.118301}
}

@article{Bhowmick2025Jun,
  title         = {Geometric and Nonequilibrium Criticality in Run-and-Tumble Particles with Competing Motility and Attraction},
  author        = {Bhowmick, A. and Mitra, S. and Mohanty, P. K.},
  journal       = {arXiv},
  year          = {2025},
  month         = jun,
  url           = {https://arxiv.org/pdf/2506.05264},
  eprint        = {2506.05264},
  archiveprefix = {arXiv},
  primaryclass  = {astro-ph.IM}
}

@article{Gutierrez2021Oct,
  title         = {Collective motion of run-and-tumble repulsive and attractive particles in one-dimensional systems},
  author        = {Gutiérrez, C. M. B. and Vanhille-Campos, C. and Alarcón, F. and Pagonabarraga, I. and Brito, R. and Valeriani, C.},
  year          = {2021},
  month         = oct,
  journal       = {Soft Matter},
  publisher     = {Royal Society of Chemistry (RSC)},
  volume        = {17},
  number        = {46},
  pages         = {10479–10491},
  doi           = {10.1039/d1sm01006a},
  issn          = {1744-6848},
  url           = {http://dx.doi.org/10.1039/D1SM01006A}
}

@article{Pu2017May,
  title         = {Reentrant phase separation behavior of active particles with anisotropic Janus interaction},
  author        = {Pu, M. and Jiang, H. and Hou, Z.},
  year          = {2017},
  month         = may,
  journal       = {Soft Matter},
  publisher     = {Royal Society of Chemistry (RSC)},
  volume        = {13},
  number        = {22},
  pages         = {4112–4121},
  doi           = {10.1039/c7sm00519a},
  issn          = {1744-6848},
  url           = {http://dx.doi.org/10.1039/C7SM00519A}
}

@misc{SM,
  key = {SM},
  note = {See Supplemental Material at [URL] for details of the simulations, derivations and additional figures.}
}

@article{Prost2015Feb,
  title         = {Active gel physics},
  author        = {Prost, J. and Jülicher, F. and Joanny, J-F.},
  year          = {2015},
  month         = feb,
  journal       = {Nature Physics},
  publisher     = {Springer Science and Business Media LLC},
  volume        = {11},
  number        = {2},
  pages         = {111–117},
  doi           = {10.1038/nphys3224},
  issn          = {1745-2481},
  url           = {http://dx.doi.org/10.1038/nphys3224}
}

@article{Martinez_Garcia_2015Jul,
  title         = {Pattern Formation in Populations with Density-Dependent Movement and Two Interaction Scales},
  author        = {Martínez-García, Ricardo and Murgui, Clara and Hernández-García, Emilio and López, Cristóbal},
  year          = {2015},
  month         = jul,
  journal       = {PLOS ONE},
  publisher     = {Public Library of Science (PLoS)},
  volume        = {10},
  number        = {7},
  pages         = {e0132261},
  doi           = {10.1371/journal.pone.0132261},
  issn          = {1932-6203},
  url           = {http://dx.doi.org/10.1371/journal.pone.0132261},
  editor        = {Sun, Gui-Quan}
}

@article{teixeira2021single,
  title={A single active ring model with velocity self-alignment},
  author={Teixeira, Emanuel F and Fernandes, Heitor CM and Brunnet, Leonardo G},
  journal={Soft Matter},
  volume={17},
  number={24},
  pages={5991--6000},
  year={2021},
  publisher={Royal Society of Chemistry}
}

@article{ourique2022modelling,
  title={Modelling micropipette aspiration with active particles},
  author={Ourique, G and Teixeira, EF and Brunnet, LG},
  journal={Physica A: Statistical Mechanics and its Applications},
  volume={589},
  pages={126661},
  year={2022},
  publisher={Elsevier}
}

\end{document}

\title{
Supplemental Material for \\
``How Quorum Sensing Shapes Clustering in Active Matter''
%Active Particles with Quorum Sensing Exhibit Reentrant Clustering\\
%Quorum Sensing Induces Reentrant Clustering in Active Particles\\
% Reentrant Clustering in Active Particles Induced by Quorum Sensing\\
% How Quorum Sensing Affects Motility-Induced Clustering\\
% Active Particles Interacting via Contact Forces and Quorum Sensing Motility Regulation\\
% Clustering of Active Matter with Steric and Quorum Sensing Interactions\\
% Self-propelled disks with quorum-sensing motility regulation\\
% Active particles interacting via chemical and mechanical motility regulation
}
\author{L. de Souza}
\email{lucas.gabrielsouza@fisica.ufrn.br}
\affiliation{Departamento de {Física} {Teórica} e Experimental, Universidade Federal do Rio Grande do Norte. Natal 59078-970, Brazil}
\affiliation{Institut für Theoretische Physik, University of Göttingen, Friedrich-Hund-Platz 1, 37077 Göttingen, Germany}
\author{E.~F. Teixeira}
\email{teixeiraemanuel9@gmail.com}
\affiliation{Instituto de {Física}, Universidade Federal do Rio Grande do Sul. Porto Alegre 91501-970, Brazil}
\author{G.~M. Viswanathan}
\email{gandhi@fisica.ufrn.br}
\affiliation{Departamento de {Física} {Teórica} e Experimental, Universidade Federal do Rio Grande do Norte. Natal 59078-970, Brazil}
\affiliation{National Institute of Science and Technology of Complex Systems, Federal University of Rio Grande do Norte. Natal 59078-900, Brazil}
\author{P. Sollich}
\email{peter.sollich@uni-goettingen.de}
\affiliation{Institut für Theoretische Physik, University of Göttingen, Friedrich-Hund-Platz 1, 37077 Göttingen, Germany}%
\affiliation{Department of Mathematics, {King’s} College London. London WC2R 2LS, UK}
\author{P.~de Castro}
\email{pablo.castro@unesp.br}
\affiliation{ICTP South American Institute for Fundamental Research, Instituto de {Física}
{Teórica}, Universidade Estadual Paulista– UNESP, São Paulo, Brazil}
\date{\today}

\maketitle

This Supplementary Material contains details on the simulations performed, the kinetic theory derivation, and the description of the supplementary movies.
We also compare simulations starting with different initial conditions to show that they converge to the same final state only on very long timescales. Moreover, we present results for extremely high QS strength.
% and the limits where the reentrant behavior vanishes

\section{\label{appen:simulation_details} Simulation details}
%\cmPa{We should probably give only the \textit{additional} simulation details. And there is no need to show the Euler-Maruyama method.}
%The system dynamics is simulated using the Euler-Maruyama method, with the equations of motion of each particle given by
% \be
% \mathbf{r}_i(t + \Delta t) = \mathbf{r}_i(t) + \left( v[ \tilde{\rho}(\mathbf{r}_i) ] \, {\bf \hat{e}}_i + \mu \displaystyle\sum_{j \neq i} \mathbf{F}_{i j} \right) \Delta t
% + \sqrt{2 \DT \Delta t} \,\boldsymbol{\xi}_i(t),
% \ee
% \be
% \theta_i(t+ \Delta t) = \theta_i(t) + \sqrt{2 \DR \Delta t} \, \eta_i(t).
% \ee
% Here, $\boldsymbol{\xi}$ and $\eta$ are normally distributed white noise.
%
% The initial condition is either homogeneous or a single compact round cluster with particles at the interface pointing towards the center of the cluster. 
% Throughout the simulations, we used ${N = 2 \times 10^4}$ particles, giving a packing fraction $\phi=0.4$, equivalent to an overall number density $\rhoglobal \approx 0.49$.
% The particles' sizes are sorted to have an equal number of large and small particles, thereby avoiding artificial clustering due to crystallization.
% We set the ratio between the two diameters to be $1.4$, with their mean equal to $\sigma=1$~\cite{Tong2019Dec}.
%The system has a packing fraction $\phi=0.4$, equivalent to an overall density $\rhoglobal \approx 0.49$.
%
Throughout the simulations, we set $\vmin = 10^{-4} \vmax$, where $\vmax=1$, and the translational diffusion constant is fixed at $\DT=5 \times 10^{-6}$.
%These small values are employed to prevent complete (and trivial) dynamical arrest in the regime of high QS strength.
In this regime, kinetic arrest effects are present, but thermal fluctuations become relevant once we get to very long timescales.
%Those values are set to prevent gels from vanishing.
%As discussed in Sec.~\ref{appen:strong_QS}, this also leads to a second reentrance in Region III (homogeneous phase).
%
The repulsion elastic constant $k$ is chosen such that particles overlap at most $3\%$ of $\sigma$ under the action of the self-propulsion forces.
Explicitly, we set ${\mu k \times (0.03\sigma) = \vmax}$, corresponding to $\mu k \approx 30 \vmax/\sigma$~\cite{Fily2012Jun}.
The QS parameters are $\RQS = 1.5 \sigma$, $\bar{\rho}=0.1$, and $\delta \rho = 0.05$. We set $\bar{\rho}$ so that a single particle within the QS range is sufficient to activate motility reduction.
Notice that our value of $\RQS=1.5 \sigma$ is not so large.
This agrees with experiments showing that, at moderate densities, the QS range is reduced due to absorption of signaling molecules.~\cite{DalCo2020Mar,Gordon1993Jun}.
%
% Consequently, the average density in Eq.~\eqref{eq:averagedensity} provides a characteristic density $K_0 = K(\sigma \, \hat{\mathbf{r}}_{ij}) \approx 0.1575\sigma^{-2}$, representing the density contribution of one particle touching another.
% Hence, $\bar{\rho}=0.635 K_0$, and $\delta \rho = \bar{\rho}/2$.

% The final time is chosen such that the system is in a stationary state for any set of parameters of interest.
We employ the Euler-Maruyama method with a time step calculated by considering an extreme interaction case. 
In the absence of quorum sensing, we require that a particle will move by at most $0.1 \sigma$ when interacting with two particles overlapping with it by the same length.
This results in the minimum time step 
$\Delta t_{\rm min} = \frac{0.1\sigma}{\vmax + 2 \mu  k \times(0.1\sigma)}$.
Rounding up $\Delta t_{\rm min}$ for our parameters, we obtain $\Delta t=0.01$.

The distance-weighted density detected by each particle is defined as
$
{\tilde{\rho}(\mathbf{r}) = \int K(\mathbf{r}-\mathbf{r}') \rho(\mathbf{r}') d^2 \mathbf{r}',}
$
where we assume each particle is a point source,
$
\rho(\mathbf{r}) = \sum_{i=1}^N \delta(\mathbf{r} -\mathbf{r}_i)
$.
Within the interaction range $R_\text{QS}$, each particle is weighted by the interparticle distance using the kernel~\cite{O'Byrne2021Dec}
\be \label{eq:kernel}
K(\mathbf{r}) = Z^{-1} \exp \left( -\frac{1}{ 1 -(r/\RQS)^2 } \right) \Theta(1 - r/\RQS),
\ee
where $Z=0.1485 \pi \RQS^2$.
Hence, for the $i$th particle, the detected density is
\be \label{eq:averagedensity}
{\tilde{\rho}(\mathbf{r}_i) = \sum_{j \neq i} K(\mathbf{r}_i-\mathbf{r}_j).}
\ee

Finally, the mean value of $\fc$ was computed from the final $30\%$ of the time series.

%%%%%%%%%%%%%%%%%%%%%%%%%%%

\section{\label{appen:longtime_bistability} Extremely long timescales}
\begin{figure*}
    \centering
    \includegraphics[width=0.6\linewidth]{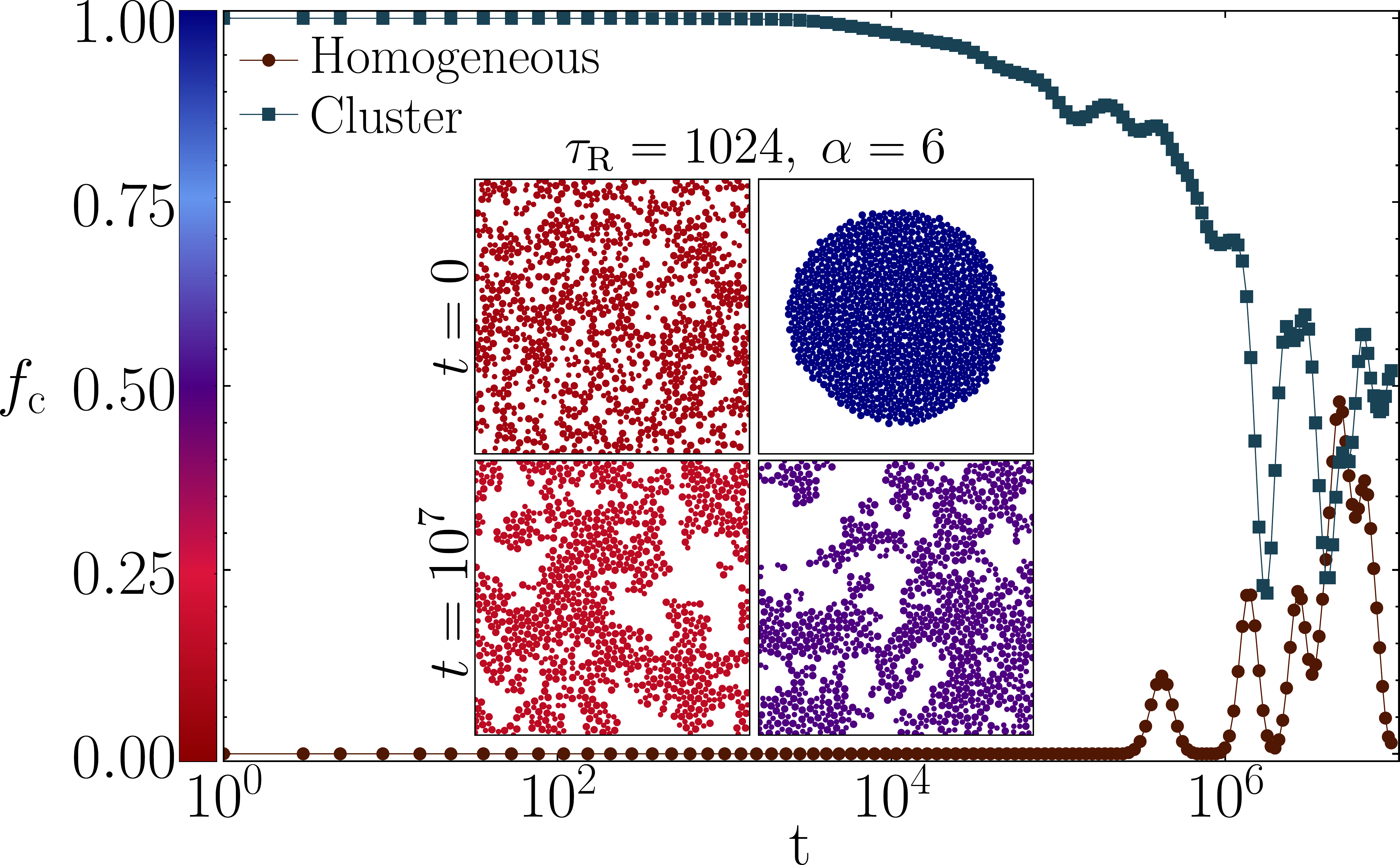}
    \caption{\justifying
    Different initial conditions converge to similar final states over long relaxation timescales.
    Evolution of the system with $N=10^3$, for homogeneous and single-cluster initial conditions. Here, $\tauR=1024$ and $\alpha=6$ (Region III in Fig.~1d).
    In the main plot, we show the time series of $\fc$, smoothed using a Gaussian filter with standard deviation equal to the time separation between two sequential data points.
    The color bar denotes the $\fc$ mean value.
    % The colorbar denotes the $\fc$ averaged from $t=7 \times 10^6$ to $t=10^7$.
    Inset shows snapshots. We note that $\fc$ for both initial conditions converges to similar values, considering fluctuations.
    The final states are nearly homogeneous phases, with some degree of clustering.
    The minimum speed $\vmin$ and the thermal diffusion help particles to escape from the cluster interface, eventually leading to the evaporation of the single-cluster into an amorphous clustering state.
    For $t \approx 10^7$, the system loses memory of the initial condition. 
    %A system with a larger $N$ requires longer relaxation timescales. % ($N = 2 \times 10^4$ in the main text)
    }
    \label{fig:long_timescales}
\end{figure*}
%
%As discussed in the main text, the effective equilibria are caused by the kinetic arrest, which can be further induced by having an initially clusterized configuration. 
Fig.~\ref{fig:long_timescales} shows the system evolution over long relaxation timescales for the homogeneous and clustered initial conditions.
Starting from a homogeneous configuration in parameter Region III, particles end up colliding with each other very soon and forming small stable clusters, which persist for long timescales due to high persistence time and QS motility reduction.
This results in a nearly homogeneous phase, with a small degree of clustering.
If the system starts from a single-cluster initial condition, the same proceeds, albeit over much longer timescales.
As particles eventually escape from the cluster interface, they interact in the gas, forming again small, long-lived clusters.
Progressively, more particles escape, ultimately leading to the complete evaporation of the initial cluster. %, causing a nearly homogeneous phase.
% On the other hand, Fig.~\ref{fig:long_timescales} shows that within Region I, the initial condition plays no role in the final level of clustering.
% The system consistently achieves the same level of clustering.
%
%We observe that, due to kinetic arrest, the initial single cluster is evaapo only after extremely long relaxation timescales, leading to the final proper degree of clustering.
These relaxation timescales grow with the number of particles. 

\section{\label{appen:kout} Details of the kinetic balance theory}
Here, we give details of the derivation of our kinetic theory for QS particles, which is based on Ref.~\cite{Redner2013Jan}.
For completeness, we provide even those details that do not change due to QS.

We assume a large single closed-packed cluster. First, we calculate the rate of particle absorption per length of the interface, $\kin$.
For particles moving in the gas phase with orientation $\theta$, $\kin(\theta)$ should be proportional to the gas density $\rhog$ and the particle speed component in the $x$-axis.
Differently from Ref.~\cite{Redner2013Jan}, our particle speed depends on the density detected via QS, $\rhog^{\rm QS}$, where $\rhog^{\rm QS}=\rhog - 1/\AQS$, as discussed in the main text. Hence, $\kin(\theta)=\rhog v(\rhog^{\rm QS}) \cos{\theta}$.
To calculate the angular average of $\kin(\theta)$ that gives the total cluster influx per unit length, we first divide by $2\pi$ to normalize and then integrate from $-\pi/2$ to $\pi/2$. This leads to
\be
\kin = \frac{1}{\pi} \rhog v(\rhog^{\rm QS}).
\ee

The total particle escape rate per unit length, $\kout$, can be written as
\be \label{eq:eq:def_kout}
\kout = - \left.\frac{1}{\sigma} \del_t \int_{-\eta}^\eta P(\theta, t) d \theta \right|_{t=0},
\ee
where $P(\theta, t)$ is the particle orientation distribution on the cluster interface, which evolves according to the rotational diffusion equation
\be
\del_t P(\theta,t) = \tauR^{-1} \del^2_\theta P(\theta,t).
\ee
We have absorbing boundary conditions ${P(\pm \eta,t)=0}$, where $\eta$ is the complementary angle of the kinetic escape cone discussed in the main text. % and ${P(\theta,0)=\frac{1}{2} \cos(\theta)}$.
Its general solution is given by
$
P(\theta, t) = \sum_{n=1}^{\infty} A_n \cos \left(\frac{\pi n \theta}{2 \eta}\right) e^{- \frac{n^2 \pi^2}{4 \eta^2} \frac{t}{\tauR}}.
$
We observe that the time decay is dominated by the first term when the system is in a steady state. 
Normalizing the solution, we find
\be \label{eq:sol_steady_state}
P(\theta, t) \approx \frac{\pi}{4 \eta} \cos \left( \frac{\pi}{2 \eta} \theta \right) \exp \left(- \frac{\pi^2}{4 \eta^2} \frac{t}{\tauR} \right).
\ee
Additionally, we need to account for the particle escape avalanches, controlled by the fitting parameter $\kappa$.
Thus, inserting Eq.~\eqref{eq:sol_steady_state} into Eq.~\eqref{eq:eq:def_kout} and multiplying by $\kappa$, we find
\be
\kout = \frac{\pi^2 \kappa}{4 \sigma \eta^2 \tauR}.
\ee

\section{\label{appen:movies} Supplementary movies}
Below, we describe the supplementary movies.
% Each movie contains a 
% %The red arrows denote the particles' orientation.
% Whenever present, 
With the exception of $\delta \rho = 0.05$, $\vmin=10^{-4} \vmax$ and $\DT=5 \times 10^{-6}$, the parameters used are indicated.
The movies can be found \href{https://drive.google.com/drive/folders/1cOeOBPvLOvU-Jl7ijQbkf9UHLHYRNJD8?usp=sharing}{here}.

\begin{enumerate}[label={\bf Movie S\arabic*:}, 
                  labelsep=0.5em, 
                  leftmargin=*, 
                  align=left, 
                  labelwidth=!, 
                  itemindent=0pt]
    \item 
    Simulation showing the competition between quorum-sensing motility reduction and persistent motion.
    Red arrows denote the particles' polarities. Particles are colored by their quorum-sensing speed; see colorbar. The orange circles show the quorum-sensing interaction region, with radius $\RQS$. Parameters:
    $\phi=0.1$, $N=50$, $\bar{\rho}=0.1$, $\tauR=16$, and $\alpha=3$.
    %If the persistence time $\tauR$ is larger than the motility reduction timescales, the particles can reach a separation larger than $\RQS$.
    
    \item
    Simulations showing evolution and steady state for the different regions of the phase diagram with homogeneous initial conditions. 
    The panels correspond (from left to right) to
    $\tauR=1,\alpha=5$ (Region I), $\tauR=32,\alpha=2.5$ (Region III), $\tauR=1024,\alpha=1$ (Region II). Particles are colored by their quorum-sensing speed; see colorbar. Parameters:
    $\phi=0.4$, $N=2 \times  10^4$, $\bar{\rho}=0.1$.
    % Formation of an active gel by quorum-sensing motility reduction.
    % $\phi=0.4$, $N=2 \times  10^4$, $\bar{\rho}=0.1$, $\tauR=4$, $\alpha=3$.
    
    \item
    Simulations showing evolution for different initial conditions. Despite long-lived transient states, both cases eventually converge to the same state for long enough relaxation timescales. Here, we use only $N=10^3$ be able to access such timescales. For $t=10^7$, the system loses memory of its initial condition as kinetic arrest fades away. Particles are colored by their quorum-sensing speed; see colorbar. The orange circles show the quorum-sensing interaction region, with radius $\RQS$.
    Parameters: $\phi=0.4$, $N=10^3$, $\bar{\rho}=0.1$, $\tauR=1024$, $\alpha=6$.

    % \item
    % Higher $\bar{\rho}$ vanishes with the reentrant behavior.
    % $\phi=0.1$, $N=50$, $\bar{\rho}=0.3$, $\tauR=16$, $\alpha=3$.
    
    % \item
    % System's behavior for extreme values of $\alpha$.
    % $\phi=0.4$, $N=10^3$, $\bar{\rho}=0.1$, $\alpha=10$.
    % For the cluster initial condition, either with $\tauR=1$ and $1024$, the cluster dissipates, leading the system to a homogeneous phase (Region III).

\end{enumerate}

%%%%%%%%%%%%%%%%%%%%%%%%%%
\bigskip
% \nocite{*}
\bibliography{QS_EV}% Produces the bibliography via BibTeX.